\documentclass[sn-nature]{sn-jnl}

\usepackage{graphicx}%
\usepackage{multirow}%
\usepackage{amsmath,amssymb,amsfonts}%
\usepackage{amsthm}%
\usepackage{mathrsfs}%
\usepackage[title]{appendix}%
\usepackage{xcolor}%
\usepackage{textcomp}%
\usepackage{manyfoot}%
\usepackage{booktabs}%
\usepackage{algorithm}%
\usepackage{algorithmicx}%
\usepackage{algpseudocode}%
\usepackage{listings}%

\newcommand{\araa}{Annu. Rev. Astron. Astrophys.}   
\newcommand{\aj}{Astron. J.}   
\newcommand{\apj}{Astrophys. J.}   
\newcommand{\apjl}{Astrophys. J. Lett.}   
\newcommand{\aap}{Astron. Astrophys.}   
\newcommand{\mnras}{Mon. Not. R. Astron. Soc.}   
\newcommand{\nat}{Nature} 
\newcommand{\na}{New Astron.}   
\newcommand{\prd}{Phys. Rev. D}   
\newcommand{\prl}{Phys. Rev. Lett.}   
\newcommand{\pasj}{Publ. Astron. Soc. Jpn}   

\theoremstyle{thmstyleone}%
\theoremstyle{thmstyletwo}%

\theoremstyle{thmstylethree}%

\begin{document}

\title[]{Evidence of a Past Merger of the Galactic Center Black Hole}


\author*[1,2]{\fnm{Yihan} \sur{Wang}}\email{yihan.wang@unlv.edu}
\author*[1,2]{\fnm{Bing} \sur{Zhang}}\email{bing.zhang@unlv.edu}

\affil*[1]{\orgdiv{Nevada Center for Astrophysics}, \orgname{University of Nevada}, \orgaddress{\street{ 4505 S. Maryland Pkwy.}, \city{Las Vegas}, \postcode{89154}, \state{NV}, \country{USA}}}

\affil[2]{\orgdiv{Department of Physics and Astronomy}, \orgname{University of Nevada}, \orgaddress{\street{505 S. Maryland Pkwy.}, \city{Las Vegas}, \postcode{89154}, \state{NV}, \country{USA}}}


\abstract{
The origin of supermassive black holes (SMBHs) residing in the centers of most galaxies remains a mystery. The Event Horizon Telescope (EHT) provided direct imaging of the SMBH Sgr A* at the Milky Way’s center, indicating it likely spins rapidly with its spin axis significantly misaligned relative to the Galactic plane’s angular momentum. Through investigating various SMBH growth models, here we show that the inferred spin properties of Sgr~A* provide evidence of a past SMBH merger. Inspired by the merger between the Milky Way and Gaia-Enceladus, which has a 4:1 mass ratio as inferred from Gaia data, we have discovered that a 4:1 major merger of SMBH with a binary angular momentum inclination angle of 145-180 degrees with respect to the line of sight (LOS), can successfully replicate the measured spin properties of Sgr A*. This merger event in our galaxy provides potential observational support for the theory of hierarchical BH mergers in the formation and growth of SMBHs. The inferred merger rate, consistent with theoretical predictions, suggests a promising detection rate of SMBH mergers for space-borne gravitational wave detectors expected to operate in 2030s.
}




\maketitle

\clearpage

The origin of SMBHs residing in the centers of most galaxies remains a profound mystery\cite{Rechstone1998,Fan2001,Lawrence2007}. Recent advancements, such as the EHT providing direct imaging of the SMBH Sgr~A* at the Milky Way’s center, have shed new light on these enigmatic objects. The EHT collaboration took the image of the Sgr~A* and provided the first physical interpretation to the data\cite{EHTI2022, EHTV2022}. This interpretation involves comparing resolved EHT data at 230 GHz and unresolved non-EHT observations from radio and X-ray wavelengths. These comparisons are made against predictions from a library of disk models based on time-dependent general relativistic magnetohydrodynamics (GRMHD) simulations, which incorporate radiative transfer based on both thermal and nonthermal electron distribution functions. In this process, 11 constraints were tested for discrete spin magnitudes: -0.94, -0.5, 0, 0.5, and 0.94; and for inclinations between the spin and the LOS of 10$^\circ$, 30$^\circ$, 50$^\circ$ 70$^\circ$ and 90$^\circ$, based on the assumption that the BH spin is aligned or anti-aligned with respect to the accretion disk. The disk models identified through these tests are the magnetically arrested disk (MAD) models surrounding a spinning Kerr BH. The `best-bet' regions in the parameter space where disk models perform well and explain nearly all the data are: a = 0.5, i = 30$^\circ$ or 150$^\circ$ and a = 0.94, i = 30$^\circ$ or 150$^\circ$. { The additional constraints from polarized images further narrow down the parameter space to a = 0.94, i = 150$^\circ$\cite{EHTVIII}}, consistent with interpretations of GRAVITY results\cite{Gravity2020b} (but see Ref. \cite{Fragione2020}).

If the spin constraints from EHT are accurate, this finding poses challenges to the accretion SMBH growth models, regardless of whether the accretion process is coherent\cite{Volonteri2007} or chaotic\cite{King2006}. In the coherent accretion model, given the disk galaxy nature of the Milky Way, the prolonged accretion disk around Sgr~A* tends to align roughly with the galaxy's angular momentum\cite{Dubois2014,Beckmann2022,Sala2024,Peirani2024} in merger-free galaxies. Therefore, if Sgr~A*'s spin originates from coherent accretion, its spin axis should not significantly misalign with the galaxy's angular momentum. Conversely, the chaotic accretion model can naturally explain large spin misalignment, as the accretion disk's orientation in this model changes with each accretion episode. Hence, if the most recent active phase of Sgr~A* has had an accretion disk almost perpendicular to the galactic plane, a misaligned spin with a large angle can be produced. However, this model struggles to account for high spin magnitude, as the random, episodic accretion pattern typically leads to an SMBH with a lower spin. Beyond accretion models, the hierarchical merger model is a promising explanation for SMBH growth\cite{Menou2001, Volonteri2003,Springel2005,Tanaka2009}. This model posits that galaxies assemble over cosmic time, leading to galaxy mergers, and ultimately, the possible merging of SMBHs at their centers. In hierarchical BH merger scenarios, a significant portion of the SMBH binary's orbital angular momentum transfers to the final spin of the merged product. As a result, SMBH mergers often yield high-spinning SMBHs. Given the arbitrary orientation of SMBH binaries, a high-spinning SMBH with substantial spin misalignment is possible. To robustly test whether the three models can reproduce the mass and spin of Sgr~A* as interpreted from EHT, we performed sophisticated simulations of SMBH growth from three different SMBH seeds\cite{Greene2020} (see Methods and Extended Data Figure~\ref{fig:BH-seeds} for details) through accretion models and merger models. We measured the spin and spin orientation with respect to the LOS when the mass of the SMBH matched the current mass of Sgr A*. We then compared the SMBH mass, spin orientation, and magnitude of our simulations with the constrained spin from EHT.

We first tested the two accretion models. Our accretion models are characterized by four parameters: the Eddington accretion ratio, $\lambda_{\rm Edd}$, indicating the efficiency of the accretion process relative to the Eddington accretion rate; the Blandford-Znajek (BZ)\cite{Blandford1977} jet power efficiency, $\eta_{\rm BZ}$, which describes the jet outflow's power in carrying away mass and angular momentum; the active accreting duty cycle fraction, $f_{\rm DC}$, detailing the SMBH's activity level; and the coherence factor, $k$, indicating the isotropy of accretion across different episodes (see Methods and Extended Data Figure~\ref{fig:mises-function}).
A zero value of $k$ implies that the accretion is completely isotropic between different episodes, while a large value of $k$ indicates that the accretion process is coherent, with the accretion disk orientation remaining nearly unchanged between episodes (see Extended Data Figure~\ref{fig:mises-function}).

\begin{figure*}
\centering
\includegraphics[width=1\columnwidth]{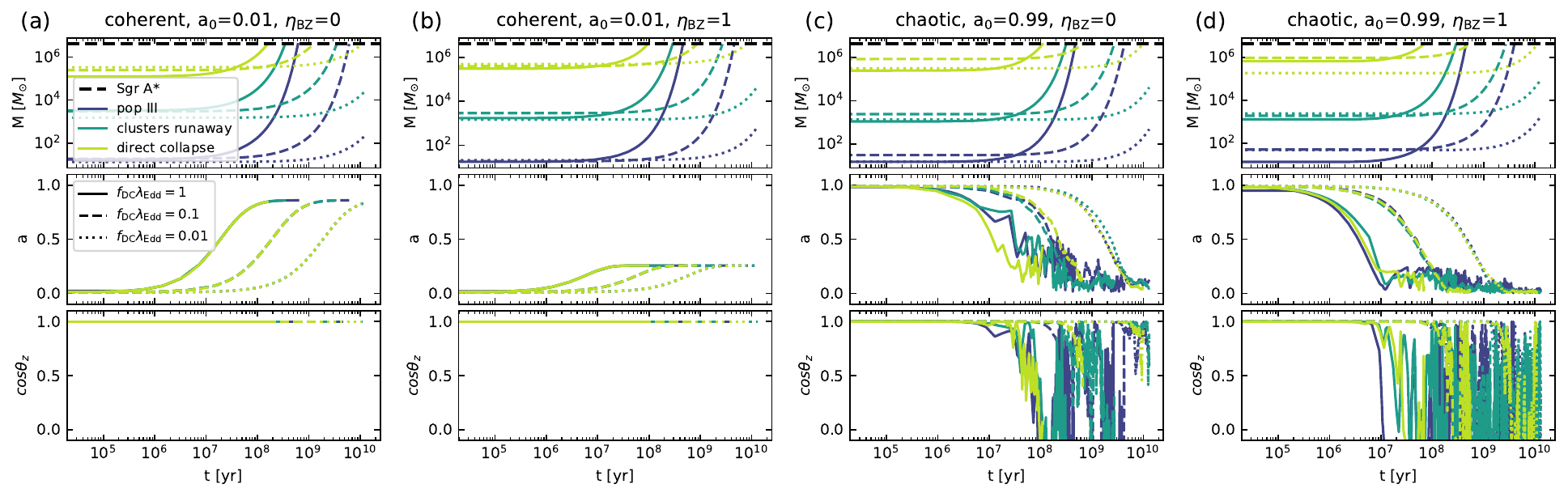}
    \caption{{\bf Examples of mass and spin evolution of coherent accretion models and chaotic accretion models from three different SMBH seeds.} Panel (a) and (c) show jet-free cases, while panel (b) and (d) show cases with magnetically saturated BZ jets.  The spin evolution is effectively independent of the SMBH seeds but shows a strong dependence on the effective accretion rate. }
    \label{fig:accretion-examples}
\end{figure*}

In panel (a) and (b) of Figure~\ref{fig:accretion-examples}, we present examples of the time evolution of the SMBH mass ($M$), spin magnitude ($\mathbf{a}=\frac{\mathbf{J}_\bullet c}{GM^2}$), and spin misalignment with respect to the galactic angular momentum ($\theta_z$) for coherent accretion models with $k=30$, originating from SMBH seeds from population-III stars, star cluster runaway collisions, and the direct collapse of gas clouds. Here, $c$ is the speed of light, $G$ the gravitational constant, and $\mathbf{J}_\bullet$ denotes the angular momentum of the SMBH. If the accretion process is jet-free, as indicated by the left panel, the SMBH seeds are spun up to approximately 0.9 within the Hubble time, regardless of whether their mass can grow to match that of Sgr A*. On the other hand, if the accretion process is associated with strong jets (which is likely), the spin of the SMBH will saturate at a certain value (see also in Ref. \cite{Lei2017}). This value depends on the jet power, where the angular momentum accreted from the accretion disk is balanced by the angular momentum carried away by the jet. Most importantly, in the coherent models, the spin of the SMBH closely aligns with the angular momentum of the galaxy throughout each accretion episode.  Conversely, panel (c) and (d) of Figure~\ref{fig:accretion-examples} illustrate examples of the time evolution for isotropic chaotic accretion models. As shown in the mass panels, the mass evolution of these SMBH seeds is very similar to that observed in the coherent accretion models. However, chaotic accretion models can significantly decrease the spin of the SMBH as the seed grows to match Sgr A*, even if the initial spin of the SMBH seeds is high and the accretion process is jet-free. As shown in those panels, While the chaotic accretion models can successfully yield a larger spin misalignment, a feature not attainable by coherent models, it falls short in replicating the high spin magnitude of Sgr~A* inferred from EHT observations.

\begin{figure*}
\centering
    \includegraphics[width=\columnwidth]{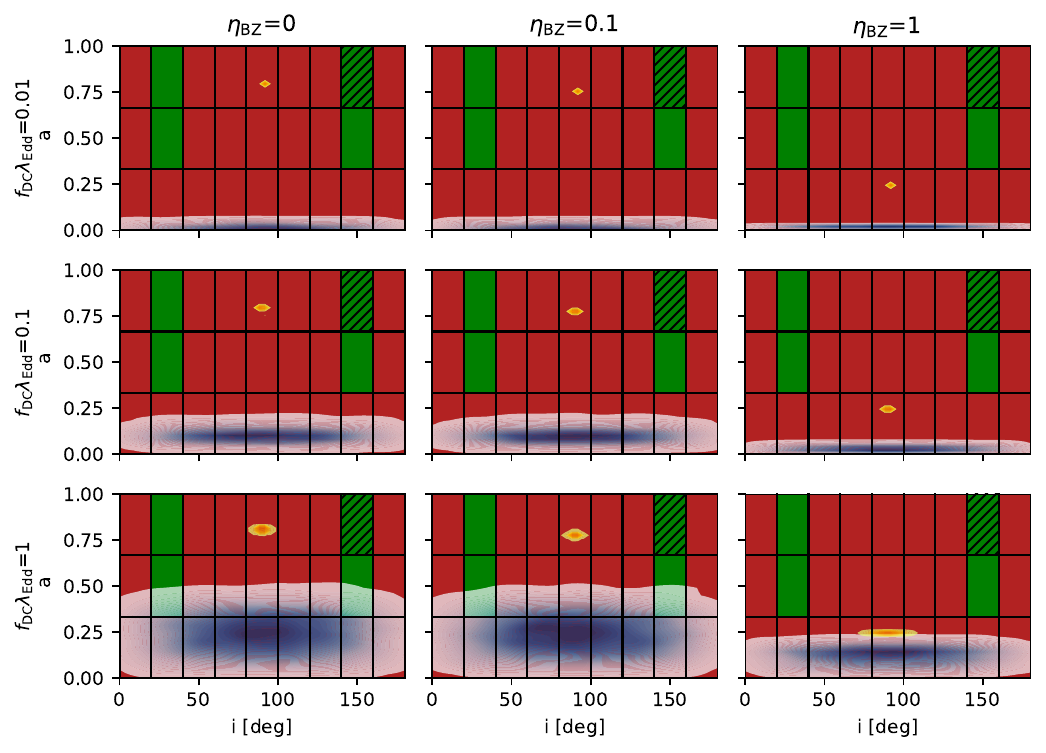}
    \caption{{\bf Kernel density estimates of the final BH spin and orientation for Sgr A*-like SMBHs accreted from direct collapse SMBH seeds.} The blue contours represent the chaotic accretion models with isotropic disk orientation, while the yellow contours represent the coherent accretion models. The left panels show the jet-free case, the middle panels show the weak BZ jet case, and the right panels show the strong BZ jet case. From top to bottom, the effective accretion rate increases. The red blocks represent regions disfavored by EHT constraints, whereas the green blocks indicate the `best-bet' regions of parameter space that perform well and explain nearly all observed data, excluding polarization. The region marked with slashes highlights the `best-bet' area, taking into account the polarization constraints.}
    \label{fig:acc-stats}
\end{figure*}

The results from Figure~\ref{fig:accretion-examples} indicate that spin evolution is effectively independent of the SMBH seed models, but is more sensitive to the effective accretion rate  $f_{\rm DC}\lambda_{\rm Edd}$ and jet power parameter $\eta_{\rm BZ}$. We conducted a suite of SMBH growth simulations across all three BH seed models, varying the accretion rates and jet powers. Figure~\ref{fig:acc-stats} displays the spin and spin-LOS angle of SMBH seeds from direct gas cloud collapses (see Extended Figures~\ref{fig:acc-stats-pop3} and ~\ref{fig:acc-stats-clust} for the other two SMBH seeds), as they grow to the mass of Sgr A* under different accretion rates and jet powers. The blue contours illustrate the isotropic chaotic accretion model, the yellow contours represent the coherent accretion model, and the green-red blocks indicate the constraints from EHT observations. Here, the red blocks represent regions disfavored by EHT constraints, while the green blocks denote the preferred parameter space where disk models perform well and explain nearly all the data. The slashed region indicates the best-bet disk models in the parameter space that passed the polarization tests as well. This figure suggests that accretion models alone are insufficient to replicate the best-bet model inferred from EHT observations.

\begin{figure*}
\centering
    \includegraphics[width=\columnwidth]{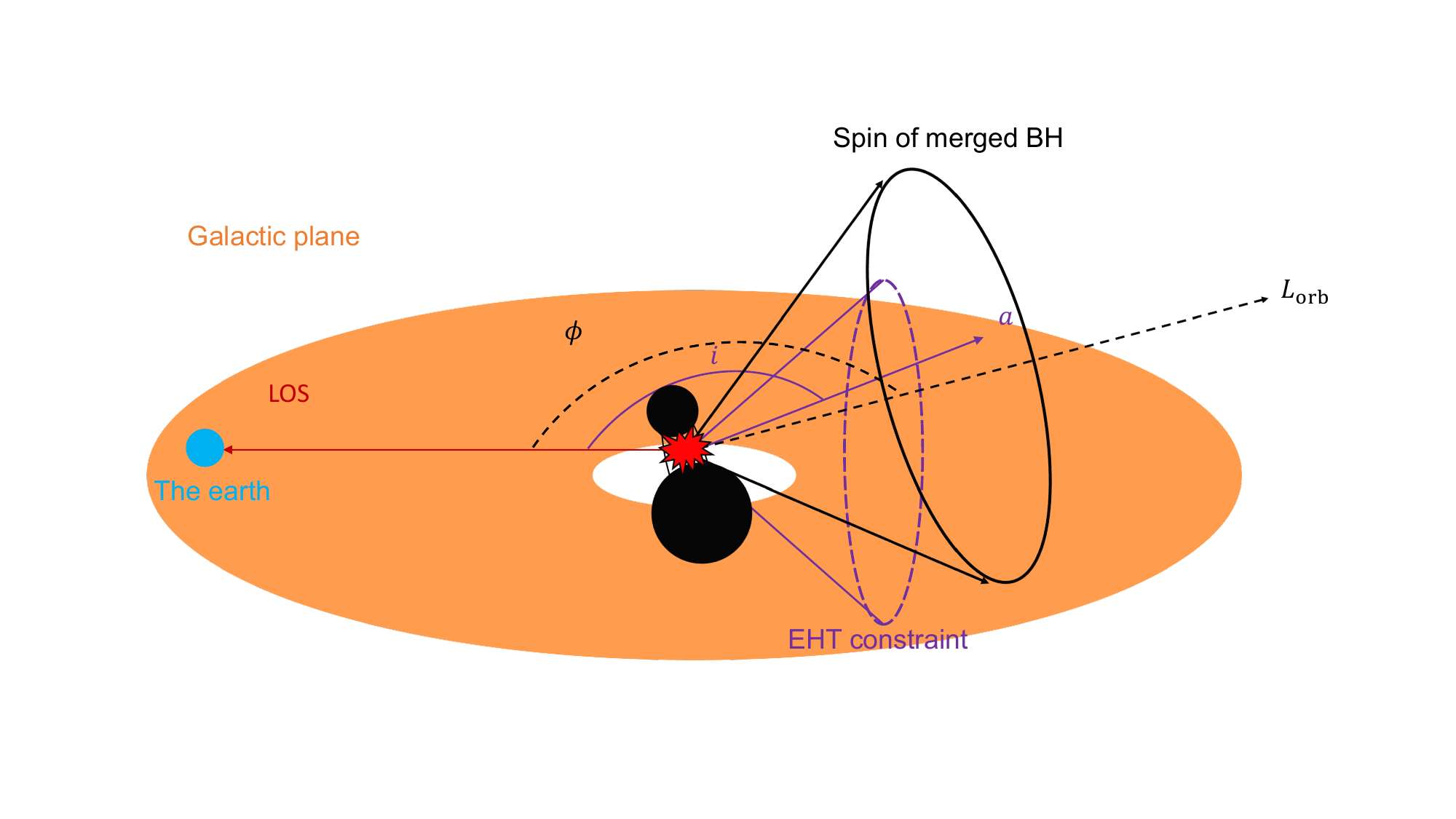}
    \caption{{\bf Schematics of SMBH merger model.} $L_{\rm orb}$ indicates the orbital angular momentum of the SMBH binary, while $\phi$ is the angle between $L_{\rm orb}$ and the line of sight (LOS) to Sgr A*. The spin after the merger that falls into the blue cone needs to satisfy the constraints from the EHT observations indicated by the purple cone. }
    \label{fig:schem}
\end{figure*}

The failure of the accretion models in reproducing the observations drive us to model the merger origin of Sgr~A*. The merger model can be parameterized by the pre-merger mass ratio between two SMBHs $q$ and their individual spin vectors $\mathbf{a}_1$ and $\mathbf{a}_2$. We use the surrogate BH merger model {\tt NRSur7dq4EmriRemnant} to predict the final spin of the merger product (see Methods). Inspired by the observed evidence of a major 4:1 galaxy merger between the Milky Way and Gaia-Enceladus approximately 10 billion years ago, as indicated by the specific chemical patterns and  motion in the nearby halo observed by the Gaia survey\cite{Helmi2018}, we fixed the mass ratio to be 4, considering the empirical near-linear correlation between the masses of the central SMBHs and their host galaxies\cite{Magorrian1998, Kormendy2013}. We explore the entire parameter space of the pre-merger spin vector $\mathbf{a}_1$ and $\mathbf{a}_2$, with arbitrary orientations and magnitude from 0.01 to 0.99. This range is chosen to encompass both the merger-free origin and the hierarchical merger origin of black holes in the merger event. Additionally, we investigate various binary SMBH orientations with respect to the LOS, denoted by $\phi$, as depicted in Figure~\ref{fig:schem}. Given the smaller mass of the secondary SMBH, we found that its spin does not significantly affect the final spin of the merger product for this 4:1 mass ratio merger (see Extended Data Figure~\ref{fig:merger-model}). Therefore, the final spin magnitude and its relative angle to the LOS are primarily determined by $\mathbf{a}_1$ and $\phi$.

\begin{figure*}
\centering
    \includegraphics[width=\columnwidth]{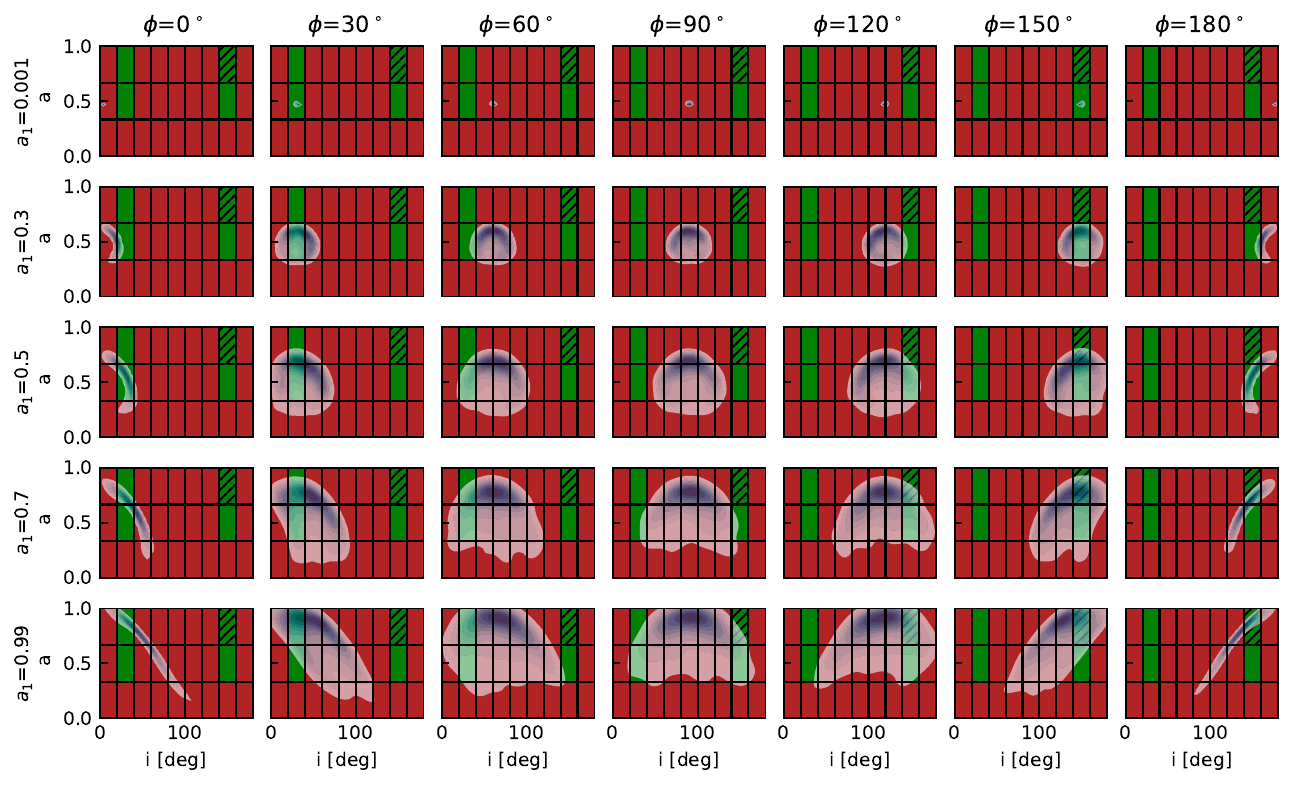}
    \caption{{\bf Kernel density estimates of the final BH spin and orientation from the 4:1 merger models.} $a_1$ and $a_2$ indicate the spin magnitudes of the primary and secondary SMBHs before the merger, respectively. The value of $\phi$ represents different binary SMBH orientations with respect to the LOS. The orientations of $a_1$ and $a_2$ are isotropically distributed to encompass both the accretion-only and merger progenitor cases. The final spin distributions show very weak dependency on $a_2$ (See Extended Figure~\ref{fig:merger-model}).}
    \label{fig:merger-stats-4}
\end{figure*}

Figure~\ref{fig:merger-stats-4} displays the kernel density estimation (KDE) of the spin resulting from simulations of the merger model, for different binary SMBH orientations $\phi$ and primary SMBH spin magnitudes $a_1$.  The orientation of $\mathbf{a}_1$ is isotropically distributed. As indicated, reproducing the best-bet model as constrained by the EHT requires a higher pre-merger spin for the primary SMBH, with $a_1$ approximately greater than 0.7 and $\phi$ in the range of $[145^\circ,180^\circ]$. No constraints on $\mathbf{a}_2$ can be established, as it does not significantly influence the spin of the merged SMBH. Additional constraints on binary SMBH orientation can be derived from findings in Ref. \cite{Helmi2018}, which reveal that a significant proportion of stellar motions in the nearby halo are in retrograde orbits. This suggests that the last major merger of the Milky Way may have been retrograde. Consequently, the angular momentum $L_{\rm orb}$ needs to be oriented below the Galactic plane. We also explored merger scenarios with various mass ratios and found that a 4:1 ratio is the most promising case for reproducing the current spin of Sgr~A*. Although subsequent minor mergers\cite{Lang2013} with specific $\mathbf{a}_1$ have not been ruled out, minor mergers with mass ratio $<$ 1/16 become less likely to reproduce the spin of the Sgr A* (See Extended Data Figure~\ref{fig:merger-stats-8} and \ref{fig:merger-stats-16}).

While the merger model can explain high-spin SMBHs with significant spin misalignment (see also in Ref. \cite{Schmitt2002}), cosmological simulations predict a relative low galaxy merger rate\cite{Springel2005, Boylan2009}. Moreover, it is unclear whether galaxy mergers efficiently lead to SMBH mergers. The uncertainty in the timescale needed to bring two SMBHs from a galactic scale (kiloparsecs) to a gravitational wave radiation scale (milliparsecs) — where they could eventually merge within the Hubble time — remains a significant challenge\cite{Milosavljevic2001, Yu2002, Sesana2015, Ryu2018}. Although we have observed some merging galaxies or even dual active galactic nuclei (AGNs), direct SMBH mergers have not been observed to date. To validate the merger fact of the Sgr~A*, we estimated the cosmological merger rate of galaxies analogous to the Milky Way, integrating data on SMBH binaries from the Millennium-II cosmological simulation\cite{Boylan2009} and accounting for the time delay to coalescence due to stellar dynamical friction, stellar scattering hardening, and gravitational wave radiation (see Methods). 

\begin{figure*}
\centering
\centering
    \includegraphics[width=\columnwidth]{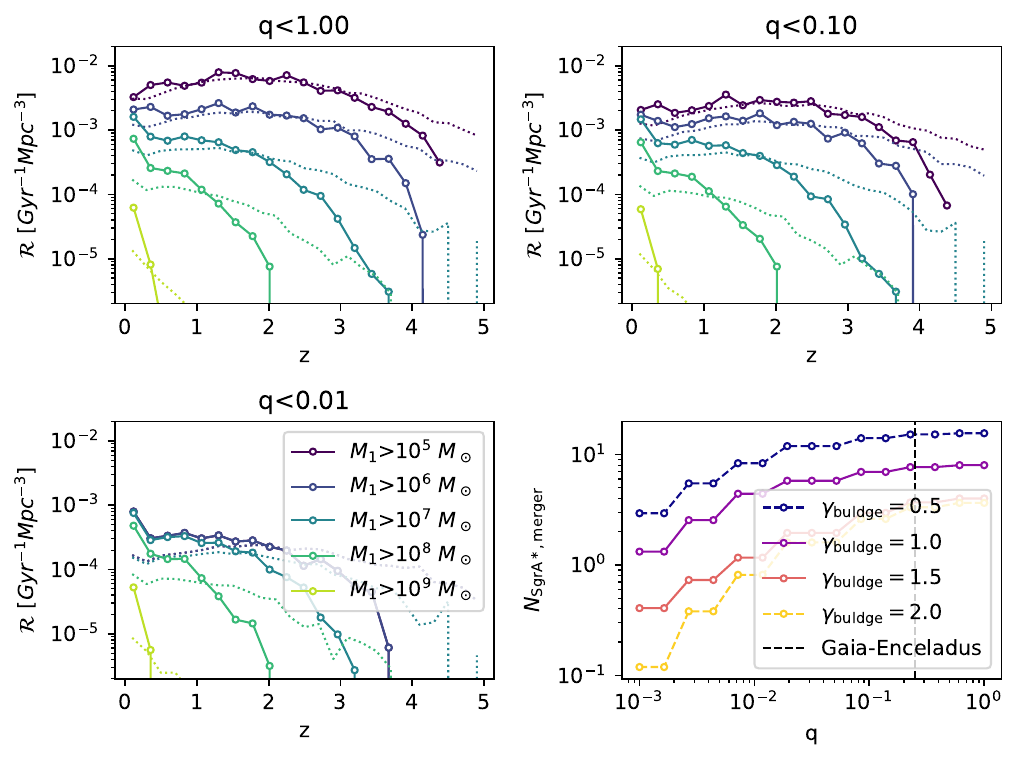}\\
    \caption{{\bf SMBH {  binary} formation and merger rates at different redshifts for different primary black hole masses and mass ratios.} Dotted lines indicate the SMBH binary formation rates while the solid lines are for SMBH merger rates. The bottom right panel displays the estimated number of past mergers experienced by the Milky Way, categorized by different merger mass ratios. { The dashed vertical line indicates the 4:1 mass ratio major mergers.} }
    \label{fig:mergerrate}
\end{figure*}

Figure~\ref{fig:mergerrate} illustrates the formation (dotted lines) and merger (solid lines) rates of SMBH binaries per volume as a function of redshift for various primary SMBH masses and mass ratios. The results indicate that major mergers peak at a redshift $z$ between 1-2, and more massive SMBHs tend to merge in the local universe. This is consistent with the hierarchical nature of galaxy formation, where more massive SMBH binaries form later. The estimated major merger rate density of SMBHs similar to Sgr~A* in the local universe is approximately $\sim 10^{-3}$ Gyr$^{-1}$ Mpc$^{-3}$, which is consistent with the results from other works\cite{Sesana2005,Volonteri2020,Li2022}. Based on this rate calculation, we estimate the number of major mergers { with mass ratio $>$ 0.25 experienced by Sgr~A* like SMBH }by adopting the galaxy number density in the local universe, which is approximately $\sim \mathcal{O}(1)$ Mpc$^{-3}$ \cite{Conselice2016}. We obtain a total number of mergers of the order of $\sim\mathcal{O}(1)$, as shown in the bottom right panel of Figure~\ref{fig:mergerrate}. This finding supports the merger origin of Sgr A*. 

We calculated the time required for Sgr A* to merge with the SMBH in Gaia-Enceladus following the galaxy merger, applying the same methodology used to estimate the SMBH merger rate from galaxy mergers. This duration is estimated to be approximately 1 billion years. Given that the merger between the Milky Way and the Gaia-Enceladus galaxy occurred roughly 10 billion years ago, we conclude that Sgr A* underwent a major merger about 9 billion years ago. This timeline is sufficiently long for the post-merger recoil of Sgr A* to settle back to the galactic center to the observed level of Brownian motion\cite{Reid2004,Ghez2008,Reid2020}, facilitated by dynamical friction\cite{Gualandris2008} (see Methods and Extended Data Figure~\ref{fig:merger-kick-4}). Additionally, this merger timeframe imposes constraints on the accretion rate following the last merger, as the spin signature imprinted on Sgr A* by the last merger has not been erased by the accretion process. We estimated a constraint of $\lambda_{\rm Edd}f_{\rm DC} \lesssim 0.5\%$ for Sgr A* following the last major merger. { The inferred accretion rate onto Sgr~A* from EHT ($<10^{-8}M_\odot$ yr$^{-1}$)\cite{EHTI2022, EHTVIII} is well below this upper limit. However, we note that the accretion rate onto Sgr~A* might have changed significantly in the past after the merger.}

The upcoming space-borne GW detectors such as the Laser Interferometer Space Antenna (LISA)\cite{LISA}, Taiji\cite{Taiji} and TianQin\cite{TianQin} are expected to detect inspiraling and merging of SMBH binaries. If the spin constraints on Sgr A* from the EHT are accurate, based on the assumption that Sgr A*'s spin is aligned with the accretion disk, it is likely that Sgr A* experienced one major merger in the past. This suggests an SMBH merger rate of approximately $\sim10^{-3}$-$10^{-2}$ Gyr$^{-1}$ Mpc$^{-3}$ up to a redshift of approximately 4. This gives an optimistic detection rate of space-borne GW detectors. Future direct detections in the 2030s will further confirm that SMBH mergers play a crucial role in the cosmological evolution of SMBHs.

\section*{Methods}\label{sec11}
Traditional theories of SMBH growth involve black hole accretion\cite{Soltan1982,Yu2002} and galaxy mergers, which lead to the mergers of SMBHs\cite{Volonteri2003,Springel2005}. In this section, we outline the models employed to simulate the growth of BHs to masses equivalent to that of Sgr~A*, incorporated with spin evolution during the growth process.

\subsection*{Seed of SMBH}

We explore the evolution of SMBH mass and spin starting from three different types of BH seeds: those originating from Population III stars, gravitational runaways in star clusters, and direct collapse. Population III stars, forming within dark matter minihalos of $10^5$-$10^6$ $M_\odot$ from primordial gas are thought to be extremely massive, ranging from 10 to 1000 solar masses \cite{Hirano2014}. Their deaths lead to stellar mass BH seeds with masses ranging from 10-100 $M_\odot$\cite{Bond1984, Madau2001}. Alternatively, the formation of intermediate-mass black holes (IMBHs), ranging from $10^3$ to $10^4$ $M_\odot$, is hypothesized to occur through stellar mergers within the dense cores of stellar clusters in metal-poor protogalaxies\cite{Bahcall1975, Begelman1978, Quinlan1990, Lee1993, Ebisuzaki2001, Miller2002, Portegies2002, Rose2022,Atallah2023}, with each cluster having a mass of approximately $10^5$ $M_\odot$ \cite{Omukai2008, Devecchi2009}. This process is facilitated by the dense gas environment, which increases the likelihood of stellar collisions. Lastly, some theories posit that the massive BH seeds can be formed from the direct collapse of clouds, resulting in massive BHs with mass ranges from $10^4$ to $10^6$ $M_\odot$\cite{Loeb1994,Begelman2006,Lodato2006}. Extended Data Figure~\ref{fig:BH-seeds} illustrates the mass distribution of the three types of BH seeds. For BHs from Population III stars, we employ stellar evolution with a Salpeter initial mass function. For gravitational runaways, we reference results from Ref. \cite{Devecchi2009}. In the case of direct collapse, we assume a log-normal distribution centered at $\log_{10}(M_{\rm BH}/M_\odot)=5$ with a standard deviation of 0.25.

\subsection*{Accretion models}\label{sec:acc}

Black hole accretion dynamics are generally classified into two types: cold flows\cite{Shakura1973,Lynden1974} , characterized by cooler, optically thick material at higher accretion rates, and hot flows\cite{Shapiro1976,Ichimaru1977}, which are hotter, optically thin, and occur at lower accretion rates. Hot flows, distinct for their lower radiative efficiency compared to the classical thin disk model, are often associated with the generation of astrophysical jets\cite{Narayan1994, Narayan1995, Blandford1999}. Such phenomena are observed in environments like low-luminosity active galactic nuclei and dormant black holes, including Sgr A*, the supermassive black hole at the center of the Milky Way.

Among various jet generation theories, the role of magnetic fields and black hole rotation is considered crucial. The Blandford-Znajek (BZ) model\cite{Blandford1977}, in particular, has gained prominence. This model suggests that the rotational energy of the black hole itself powers the jets. The BZ model's explanation aligns closely with observations and numerical simulations\cite{McKinney,Ghisellini2014, Narayan2022}, indicating that relativistic jets primarily originate from the BZ process.

In the BZ model, a large-scale poloidal magnetic field penetrates the ergosphere of a rotating black hole and intersects the event horizon. The rotation of the black hole causes frame-dragging, which in turn generates a toroidal magnetic field and a Poynting flux. The essence of the BZ process is the possibility of an inward electromagnetic energy flux within the ergosphere that appears negative when measured from infinity. This negative flux crosses the event horizon, reducing the black hole's mass-energy and angular momentum. Consequently, an outgoing jet emerges, Poynting-dominated, carrying away positive energy and angular momentum. The power of the jet in the BZ model can be described as\cite{Blandford1977},
\begin{eqnarray}
p_{\rm BZ} &\sim& 2.5\left(\frac{r_g}{r_\bullet}\right)^2 a_\bullet^2\eta_{\rm BZ}^2\dot{M}c^2,\label{eq:bz}
\end{eqnarray}
where $r_g$ and $r_\bullet$ represent the radii of Schwarzschild and Kerr black hole, respectivley, $a_\bullet=\frac{\mathbf{J}_\bullet c}{GM_\bullet^2}$ is the normalized black hole spin, and $\eta_{\rm BZ} = \Phi/\Phi_{\rm MAD}$ represents the ratio of the actual magnetic flux threading the black hole's horizon ($\Phi$) to the maximum saturated magnetic flux ($\Phi_{\rm MAD}$). A value of $\eta_{\rm BZ}=1$ signifies that the magnetic flux from the magnetized accretion flow has saturated the black hole, leading to a magnetically arrested disk (MAD)\cite{Igumenshchev2003, Narayan2003}. Conversely, a lower value of $\eta_{\rm BZ}$ suggests that the accretion disk is in a state of Standard And Normal Evolution (SANE)\cite{Narayan2012b}. { It is worth noting that Equation~\ref{eq:bz} applies for low to moderate spins. There are higher-order corrections as spin approaches 1\cite{Tchekhovskoy2011, Ricarte2023, Lowell2024}. }The angular momentum carried out by the BZ jet then can be estimated by
\begin{eqnarray}
\dot{\mathbf{J}}_{\rm BZ} &=& p_{\rm BZ}/\Omega_f \hat{a_\bullet},
\end{eqnarray}
where $\hat{a_\bullet}$ denotes the unit vector along the BH spin $a_\bullet$ vector, and
\begin{eqnarray}
\Omega_f &=& \frac{a_\bullet c}{2r_\bullet} = \frac{a_\bullet}{1+\sqrt{1-a_\bullet^2}}\frac{c}{2r_g}\,.
\end{eqnarray}
Incorporated with the standard accretion model that excludes jet contributions, the equations governing the mass and angular momentum evolution of an accreting black hole can be expressed as follows\cite{Lei2017}:
\begin{eqnarray}
\frac{dM_{\bullet}}{dt}=\dot{M}E_{\rm ms} - p_{\rm BZ}/c^2, \label{eq:dmdt}\\
\frac{d\mathbf{J_\bullet}}{dt}=\dot{M}J_{\rm ms}\hat{\mathbf{J}}_d-\dot{\mathbf{J}}_{\rm BZ}, \label{eq:dldt}
\end{eqnarray}
where $\hat{\mathbf{J}}_d$ is the unit vector along the disk angular momentum, $E_{\rm ms}$ and $J_{\rm ms}$ are the specific energy and specific angular momentum at the innermost radius $r_{\rm ISCO}$, which are defined as\cite{Page1974}
\begin{eqnarray}
E_{\rm ms} &=& \frac{4\sqrt{r_{\rm ISCO}/r_g}-3a_\bullet}{\sqrt{3}r_{\rm ISCO}/r_g}, \\
J_{\rm ms} &=&\frac{GM_\bullet}{c}\frac{6\sqrt{r_{\rm ISCO}/r_g}-4a_\bullet}{\sqrt{3}\sqrt{r_{\rm ISCO}/r_g}}. 
\end{eqnarray}
For each accretion episode, Equations~\ref{eq:dmdt} and \ref{eq:dldt} are numerically solved using given parameterized values for $\dot{M}$ and $\eta_{\rm BZ}$.

During each accretion episode, we adopt a constant accretion rate $\dot{M}=\lambda_{\rm Edd}\dot{M}_{\rm Edd}$, where $\lambda_{\rm Edd}$ signifies the accretion efficiency, and $\dot{M}_{\rm Edd}=\frac{L_{\rm Edd}}{\epsilon c^2}$ represents the Eddington accretion rate. Here, the radiation efficiency $\epsilon$ is set to be 0.1. Accretion stops when the material within the self-gravitating radius $R_{\rm sg}$ is exhausted. The accretion time for each episode, $\tau_{\rm acc}$, can be approximated by $\frac{M_d}{\dot{M}}$, with $M_d$ being the total disk mass inside the self-gravitating radius $R_{\rm sg}$. The accretion time can be parameterized as\cite{King2008},
\begin{equation}
\tau_{\rm acc}\sim \frac{M_d}{\dot{M}} = 1.12\times 10^6 \left(\frac{\alpha}{0.01}\right)^{-2/27}\left(\frac{\epsilon}{0.1}\right)^{22/27}\left(\frac{L}{0.1L_{\rm Edd}}\right)^{-22/27}M_8^{-4/27} {\rm yr}. 
\end{equation}
After an accretion episode concludes, the SMBH enters a dormant phase until the onset of the next episode. The interim period, known as the waiting time, is governed by the active galactic nucleus (AGN) duty cycle fraction $\tau_{\rm acc}/(\tau_{\rm acc}+\tau_{\rm wait})$, which is empirically observed to range from 1\% to 10\%. 

\subsection*{Coherent/Chaotic accretion}
The Lense-Thirring precession can result in the counter-alignment of a black hole's spin vector $J_\bullet$ with the angular momentum vector $J_d$ of its accretion disc\cite{King2006}. This counter-alignment process is contingent upon the condition that the angle $\theta$ between the vectors exceeds $\pi/2$, and that the angular momentum of the disc is less than twice that of the black hole, as $J_d < 2J_\bullet$.

According to the analysis, if the angular momentum of the disc is less than twice that of the black hole, $J_d < 2J_\bullet$, then for randomly oriented angular momentum vectors, the proportion $f_c$ of systems that will exhibit counter-alignment is given by the relationship:

\begin{equation}
f_c = \frac{1}{2}\left(1-\frac{J_d}{2J_\bullet}\right)
\end{equation}

In scenarios where the disc's angular momentum is less than that of the black hole $J_d <J_\bullet$, and accretion occurs through randomly oriented events, the black hole is likely to experience alternating episodes of spinup and spindown, a process referred to as chaotic accretion\cite{King2006,King2008,Chen2023}. Conversely, in situations where $J_d >J_\bullet$, the black hole is expected to consistently gain angular momentum, or spin up, which is indicative of coherent accretion. This behavior would persist even if the black hole and the disc were initially counter-aligned.

In our accretion models, we characterize the extent of chaotic accretion using the Mises distribution function for the disk angular momentum vector $\mathbf{J}_d$\cite{Bustamante2019}
\begin{equation}
p(\theta,k) = \frac{e^{k\cos\theta}}{2\pi J_0(k)},\label{eq:pk}
\end{equation}
where $J_0(k)$ is the zero-th order cylindrical Bessel function for the first kind. When $k=0$, the model yields a perfectly isotropic chaotic accretion. In contrast, a large $k>30$ results in nearly coherent accretion, where $\theta$ is consistently close to zero.

\subsection*{Merger models}
In the realm of binary black hole systems, those with black hole spins misaligned from the orbital angular momentum present significant challenges for analytical or semi-analytical modeling. The complex interactions between the spins, the orbital angular momentum, and each other lead to the precession of the system around the total angular momentum vector. To predict the final spin of the merged black hole, several faster approximate models  have been developed\cite{Varma2019b, Boschini2023}, covering waveforms\cite{Blackman2017} and remnant properties of merging black holes\cite{Varma2019}. Those surrogate models for precessing binary black hole with generic spins and unequal masses are based on large sample of numerical relativity simulations, which are proven to be accurate in predicting the final spin magnitude and orientation with relative error at the order of $10^{-2}$ with 90th percentiles confidence\cite{Boschini2023}. With those surrogate models, we explore the probability that the SgrA* is the merger product of two SMBHs. We have adopted the latest {\tt NRSur7dq4EmriRemnant} model\cite{Boschini2023}, which is based on a seven-dimensional parameter space characterizing generically precessing binary black hole systems. This model incorporates the binary black hole mass ratio $q$, along with the two pre-merger spins $\mathbf{a}_{\bullet 1}$ and $\mathbf{a}_{\bullet 2}$.

In addition to the seven parameters described by the {\tt NRSur7dq4EmriRemnant} model, our SMBH merger model includes an additional parameter, $\phi$, which represents the inclination angle of the SMBH binary angular momentum with respect to the LOS, as illustrated in Fig.~\ref{fig:schem}. We explore 9 different inclinations ranging from $0^\circ$ to $180^\circ$, with 20$^\circ$ equally spaced intervals. The spin magnitudes and orientations of $\mathbf{a}_{1}$ and $\mathbf{a}_{2}$ (with respect to the SMBH binary angular momentum) are randomly distributed, with magnitudes ranging from 0 to 1 to cover all possible pre-merger origins.

Recent studies of stellar surveys show specific chemical patterns and a retrograde motion in the nearby halo, suggesting a major past accretion event in the Galaxy. It has been found that the inner halo mainly consists of debris from a system, named Gaia-Enceladus\cite{Helmi2018}, once larger than the Small Magellanic Cloud. Given the estimated mass ratio of 4:1, the merger between the Milky Way with Gaia-Enceladus around 10 billion years ago likely caused significant dynamical heating of the Galaxy's early thick disk. Based on this observation, we fix the mass ratio of the merging black holes to be 4:1.

Owing to the 4:1 mass ratio, the final galactic plane is predominantly determined by the galactic plane of the pre-merger massive galaxy, which is our Milky Way. We assume that the final galactic plane aligns with the initial galactic plane of the more massive galaxy. The final spin in our merging model is computed using the surrogate model, which takes the previously described input parameters. 

\subsection*{Cosmological merger rate estimation}
The cosmic population of SMBHBs is derived from the data of the Millennium-II N-body simulation \cite{Boylan2009}, which has the resolution to distinguish dark matter halos down to $\sim10^8 M_\odot$. This dark matter framework is augmented with semianalytic models of galaxy evolution to trace the history of galaxies down to $\sim10^6M_\odot$, which host central black holes as small as $\sim 10^4M_\odot$. All post-processing employs the same cosmological parameters as Millennium-II: $h=0.73$, $\Omega_\Lambda=0.75$, $\Omega_M=0.25$, and $\Omega_k=\Omega_r=0$. While these cosmological parameters are not the most current, variations of a few percent in $h$, $\Omega_\Lambda$, or $\Omega_M$ do not significantly affect our results.

We identify all mergers between galaxies, each containing an SMBH, totaling $N_\bullet=169435$ events throughout the Millennium-II simulation's span. { This was done by assuming an MBH occupation fraction of unity down to 10$^8M_\odot$ stellar mass halos. This assumption slightly overestimates the SMBH binary merger rate at the lower mass end, as the MBH occupation fraction in these halos can drop to approximately $\sim$40\%\cite{Tremmel2024} for 10$^8M_\odot$ stellar mass halos.} We construct a 3D grid by logarithmically binning $\log_{10}(M_1/M_\odot)$ and the mass ratio $q$ over the intervals $M_1\in[10^5,10^{10}]M_\odot$ and $q\in[10^{-4},1]$, and linearly binning the redshift $z$ from 0 to 5 in line with the redshifts in Millennium-II snapshots. The redshift of each merger is matched with the snapshot where the progenitor galaxy is detected, assuming that the dynamical friction timescale for the SMBHs to form a close binary is shorter than the average 300 Myr between simulation snapshots.

By dividing the number of SMBHBs in each bin by the co-moving volume $V_c$ of the Millennium-II simulation, we approximate the differential SMBHB formation rate as
\begin{equation*}
\frac{d^4 N_{\bullet}}{d z\,dq\,dM_1\,d V_c}\,.
\end{equation*}

For those SMBHs that become a bound binary, the SMBHs start a long journey that bring them from separations of tens of kpc down to milli-parsec scales, below which they merge through GWs. The transition into the GW domain depends on energy exchanges with low-angular-momentum stars in the galaxy's nucleus\cite{Begelman1980}, or on the removal of angular momentum via gravitational torques from a gas disk, which can diminish their separation \cite{Goldreich1980}. It might also involve a mix of these two mechanisms. Recent advancements in direct N-body simulations, Monte Carlo methods, and scattering experiments have provided a more hopeful outlook on what has been deemed the primary bottleneck in binary evolution for nearly four decades: the ``final parsec problem". This issue, described as the depletion of low-angular-momentum stars\cite{Begelman1980}, seems less problematic. Findings suggest that the evolution of SMBHBs through stellar scattering potentially lead to a merger within less than approximately 1 billion years\cite{ Berczik2006, Sesana2015,Vasiliev2015}, especially when factors such as rotation, the triaxial shape of galaxies, and the granularity of stellar distribution are considered. SMBHB hardening through stellar scattering from pairing to the final coalescence can be described by\cite{Sesana2015}
\begin{eqnarray}
\frac{da_{\rm b}}{dt} &=& \frac{da_{\rm b}}{dt}\bigg|_{3b}+\frac{da_{\rm b}}{dt}\bigg|_{\rm GW}\\
&=& -\frac{GH\rho_{\rm inf}}{\sigma_{\rm inf}}a^2_{\rm b} -\frac{64G^3M_1M_2(M_1+M_2)f(e)}{5a_{\rm b}^3c^5}, \nonumber
\end{eqnarray}
with
\begin{equation}
f(e) = \frac{1+(73/24)e^2+(37/96)e^4}{(1-e^2)^{-7/2}}, 
\end{equation}
where the first term represents the rate of binary hardening due to stellar scattering\cite{Quinlan1996, Sesana2006}, characterized by the dimensionless hardening rate $H$, stellar density $\rho_{\rm inf}$ and velocity dispersion $\sigma_{\rm inf}$ at the binary influence radius $r_{\rm inf}$. The influence radius is defined as the radius within which the enclosed stellar mass is twice the mass of the primary SMBH. The second term captures the effects of GW radiation\cite{Peters1963,Peters1964}, which is characterized by the masses of the primary and secondary SMBHs in the binary ($M_1$ and $M_2$, respectively) and the binary's eccentricity $e$. These terms together describe the evolution of the separation between the SMBHs under the combined influences of dynamical interactions with stars and energy loss due to GW radiation. The SMBHBs spend most of their lifetimes at the transition separation where $da_b/dt|_{3b}=da_b/dt|_{\rm GW}$, which is
\begin{equation}
a_{\rm GW} = \left(\frac{64G^2\sigma_{\rm inf} M_1M_2(M_1+M_2)f(e)}{5c^5H\rho_{\rm inf}}\right)^{1/5}. 
\end{equation}
The corresponding timescale at this separation is
\begin{equation}
\tau_{\rm delay} \sim \frac{\sigma_{\rm inf}}{GH\rho_{\rm inf} a_{\rm GW}},
\end{equation}
which can be used to estimate the time from SMBHB formation to final coalescence. The dimentionless hardening rate $H$ is estimated to be\cite{Sesana2015} $\sim 10$, and the velocity dispersion $\sigma_{\rm inf}$ can be estimated from the M-$\sigma$ relation \cite{Kormendy2013}
\begin{eqnarray}
M_9 &=& 0.309 (\frac{\sigma_{\rm inf}}{200 {\rm km\cdot s^{-1}}})^{4.38}, 
\end{eqnarray}
where $M_9 = M_\bullet/(10^9M_\odot)$ and the stellar density at the influence radius can be obtained from the nuclei stellar cluster model\cite{Dehnen1993}
\begin{equation}
\rho(r) =\frac{(3-\gamma)M_*}{4\pi}\frac{r_0}{r^\gamma(r+r_0)^{4-\gamma}}, 
\end{equation}
where $M_*\sim (1.84M_9^{0.86}\times 10^{11}) M_\odot$ denotes the total mass of the stars in the bulge and $\gamma$ represents the inner cusp slope. The influence radius can be obtain by equations where\cite{Dehnen1993,Dabringhausen2008},
\begin{eqnarray}
r_{\rm inf} &=& \frac{r_0}{(\frac{M_*}{2M_\bullet})^{1/(3-\gamma)} -1 }, \\
r_0 &=& \frac{4}{3}R_{\rm eff} ( 2^{1/(3-\gamma)}-1),\\
R_{\rm eff}/{\rm pc} &=& \left\{ 
\begin{array}{ll}
    {\rm max}(2.95 M_{*,6}^{0.596},34.8M_{*,6}^{0.399}),\quad & {\quad \rm elliptical\quad galaxies}, \\
    2.95 M_{*,6}^{0.596},\quad &{\quad \rm spiral\quad galaxies}.
\end{array}
\right.
\end{eqnarray}

For each pair of SMBHs derived from the Millennium-II simulation, we can estimate the coalescence time starting from binary formation and calculate the redshift at which they ultimately merge using the following equation
\begin{eqnarray}
\tau_{\rm delay} &=& \int_{z^\prime}^z \frac{d t}{d \bar{z}}d \bar{z}\nonumber\\
&=& \frac{1}{H_0}\int_{z^\prime}^z\frac{d \bar{z}}{(1+\bar{z})\sqrt{\Omega_r(1+\bar{z})^4+\Omega_M(1+\bar{z})^3+\Omega_k(1+\bar{z})^2+\Omega_\Lambda}}\,,
\end{eqnarray}
where $z$ denotes the redshift at which the SMBHB forms, and $z^\prime$ represents the redshift at which the SMBHB coalesces. If the coalescence delay time, $\tau_{\rm delay}$, is sufficiently long, it could result in an unphysical, negative $z^\prime$. This indicates that the SMBHB is still in the hardening phase. Such shrinking/stalled SMBHBs are excluded from our dataset. Then the differential SMBHB merger rate can be obtained as
\begin{equation}
\mathcal{R}_{\rm merger}(z^\prime,q,M_1) = \frac{d^4 N_{\rm merger}}{d V_c\,d q\,d M_1\,d t}\bigg|_{z^\prime>0}\,.
\end{equation}

\subsection*{Recoil velocity from SMBH merger}
The emission of gravitational waves during the merger of two SMBHs imparts a recoil, or 'kick', to the resulting SMBH\cite{Bekenstein1973,Blecha2008}. If Sgr~A* underwent a major merger in the past, the recoil velocity from this merger would cause Sgr~A* to follow an elliptical orbit. After receiving this kick, the movement of the SMBH undergoes three distinct phases\cite{Gualandris2008}. Initially, the SMBH oscillates back and forth with decreasing amplitude, losing energy through dynamical friction with each pass through the galactic core. This motion can be precisely modeled using dynamical friction. As the magnitude of motion decreases to about the core radius size, the SMBH and the core begin to oscillate around their common center of gravity, with these oscillations diminishing more slowly than the initial dynamical friction phase. In the final phase, the SMBH stabilizes and reaches thermal equilibrium with the surrounding stars. For the Milky Way, the stellar velocity dispersion at the Galactic center is approximately 100 km/s, leading to an equilibrium state of Brownian motion for Sgr~A* at about 0.2 km/s\cite{Merritt2007}. Observations of Sgr~A*'s peculiar motion perpendicular to the galactic plane\cite{Reid2004,Ghez2008,Reid2020} are consistent with this prediction, suggesting that Sgr~A* is currently positioned quietly at the center of the Milky Way. However, the peculiar motion in the galactic plane is measured to be above the Brownian motion velocity. Therefore, the existence of a smaller BH or IMBH\cite{Hansen2003, Maillard2004, Gurkan2005, Gualandris2009, Chen2013, Generozov2020, Naoz2020, Zheng2020, RoseNaoz2022, Will2023} has not been completely ruled out. The most recent study posits a strong mass constraint on the secondary BH, indicating it to be less than 2000 $M_\odot$\cite{Gravity2023}. 

We employ the {\tt NRSur7dq4Remnant} model to calculate the kick velocity of the final SMBH in our merger scenarios, as illustrated in the left panel of Extended Data Figure~\ref{fig:merger-kick-4}. Given that a 4:1 major merger with a spin parameter greater than or equal to 0.7 can accurately reproduce the spin of Sgr~A*, it is likely that Sgr~A* would receive a substantial kick velocity, ranging from hundreds to thousands of kilometers per second. { By interpolating the results of Ref. \cite{Gualandris2008}(model A1 in their figure 15),} we calculate the time required for Sgr~A* to settle into the Brownian motion phase, with peculiar motion velocity $\sim$ 0.2 km/s. As demonstrated in the right panel of Extended Data Figure~\ref{fig:merger-kick-4}, for the entire parameter regime we explored, the recoil velocity Sgr~A* acquired from the major merger could be sufficiently dissipated by the stars within the Milky Way's core over the 9 billion years following the merger, consistent with the observations.

\backmatter


\subsection*{Data availability}
\noindent

\noindent
Data are available at \url{https://figshare.com/articles/dataset/Sgr_A_data/26112379}.

\subsection*{Code availability}
\noindent
The code and data process script are available at \url{https://github.com/YihanWangAstro/Sgr-A-PubCode}.

\section*{Acknowledgements}
Y.W. and B.Z. acknowledge the support from NASA 80NSSC23M0104 and the Nevada Center for Astrophysics. Y.W. acknowledges useful discussions with Tamara Bogdanovic regarding the recoil velocity of major mergers, with Douglas N. C. Lin on subsequent accretions following the merger, as well as with Barry McKernan and Daniel Stern concerning coherent accretions.

\section*{Author Contributions} 
B.Z. proposed the idea of this paper; Y.W. developed the theoretical models; Y.W. and B.Z. analyzed the results and discussed the theoretical models. All authors contributed to the analysis or interpretation of the data and to the final version of the manuscript.

\section*{Competing Interests} 
The authors declare no competing interests.

\section*{Tables}

\section*{Figure Legends/Captions }\label{sec6}

\setcounter{figure}{0} 
\clearpage

\section*{Extended Figure Legends/Captions }

\begin{figure*}[ht]
\centering
\includegraphics[width=\columnwidth]{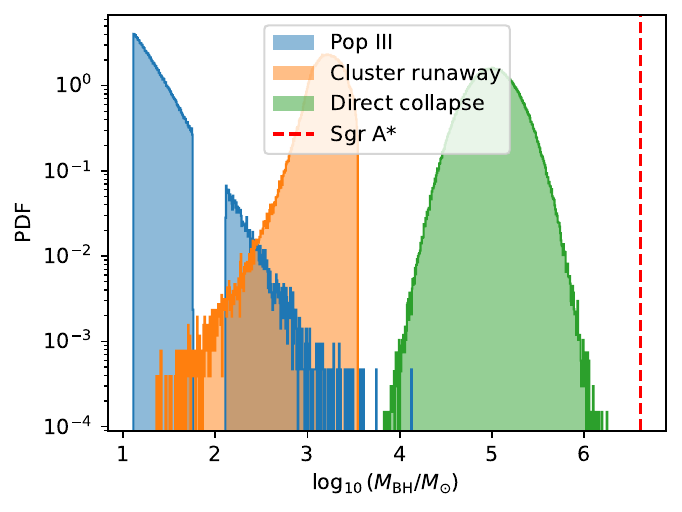}
\caption{\textbf{SMBH seeds mass distribution from three different formation channels.} This figure presents the mass distribution of SMBH seeds derived from three distinct formation channels. The blue distribution represents SMBH seeds formed from population-III stars. The orange distribution illustrates the SMBH seeds resulting from cluster runaway collisions. The green distribution shows the mass distribution from direct collapse of clouds and dark matter. }
\label{fig:BH-seeds}
\end{figure*}

\begin{figure*}
\centering
\includegraphics[width=\columnwidth]{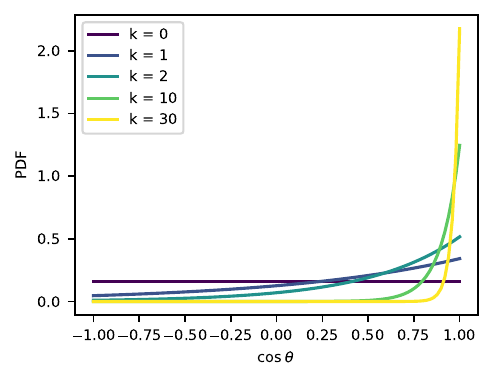}
\caption{{\bf Distribution of the accretion disk misalignment angle for different values of $k$ in Mises function.} $k=0$ indicates isotropically distributed accretion model, while large value of $k$ ($\sim>$30) is asymptotic to coherent accretion model. }
\label{fig:mises-function}
\end{figure*}

\begin{figure*}
\centering
    \includegraphics[width=\columnwidth]{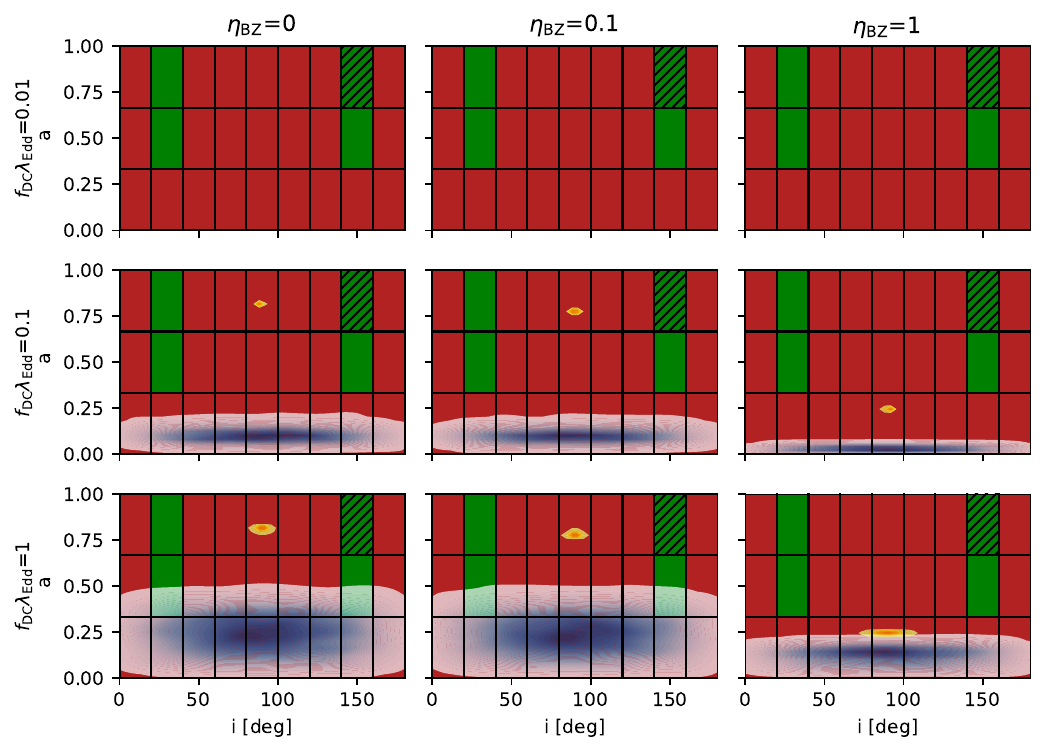}
    \caption{{\bf Kernel density estimates of the final BH spin and orientation for Sgr A*-like SMBHs accreted from Pop-III SMBH seeds.} The blue contours represent the chaotic accretion models with isotropic disk orientation, while the yellow contours represent the coherent accretion models. The left panels show the jet-free case, the middle panels show the weak BZ jet case, and the right panels show the strong BZ jet case. From top to bottom, the effective accretion rate increases. The red blocks represent regions disfavored by EHT constraints, whereas the green blocks indicate the `best-bet' regions of parameter space that perform well and explain nearly all observed data, excluding polarization. The region marked with slashes highlights the `best-bet' area, taking into account the polarization constraints. The absence of contours indicates failures to accrete to Sgr A* mass within the Hubble time.}
    \label{fig:acc-stats-pop3}
\end{figure*}

\begin{figure*}
\centering
    \includegraphics[width=\columnwidth]{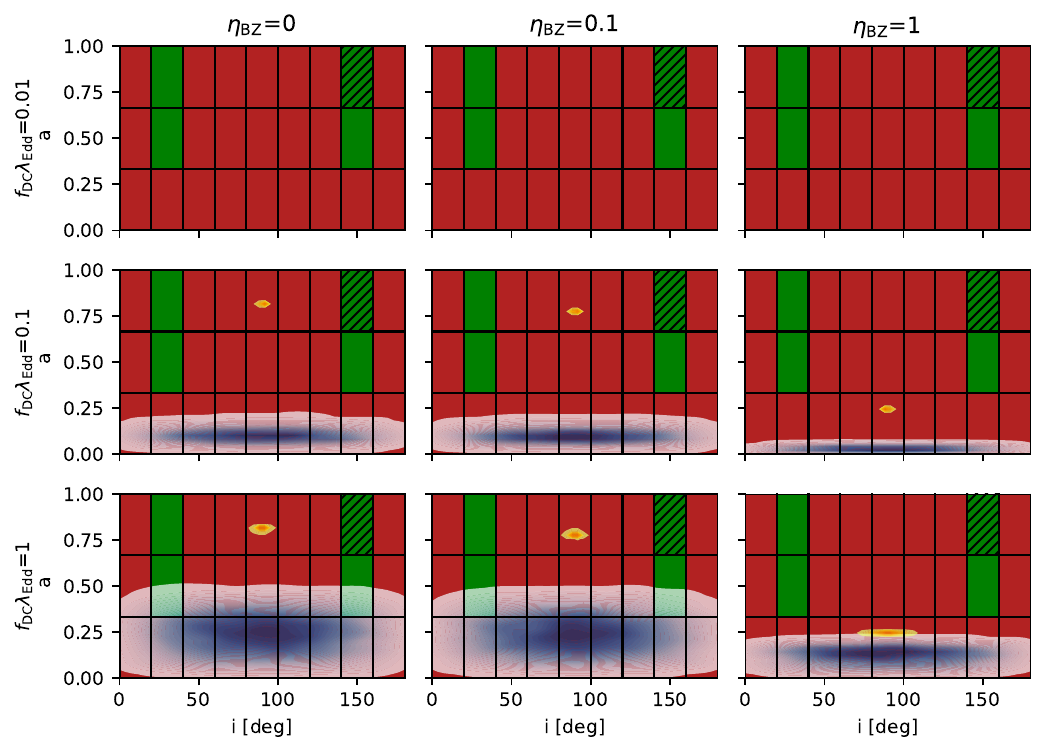}
    \caption{{\bf Kernel density estimates of the final BH spin and orientation for Sgr A*-like SMBHs accreted from star cluster runaway SMBH seeds.} The blue contours represent the chaotic accretion models with isotropic disk orientation, while the yellow contours represent the coherent accretion models. The left panels show the jet-free case, the middle panels show the weak BZ jet case, and the right panels show the strong BZ jet case. From top to bottom, the effective accretion rate increases. The red blocks represent regions disfavored by EHT constraints, whereas the green blocks indicate the `best-bet' regions of parameter space that perform well and explain nearly all observed data, excluding polarization. The region marked with slashes highlights the `best-bet' area, taking into account the polarization constraints. }
    \label{fig:acc-stats-clust}
\end{figure*}

\begin{figure*}
\centering
\includegraphics[width=\columnwidth]{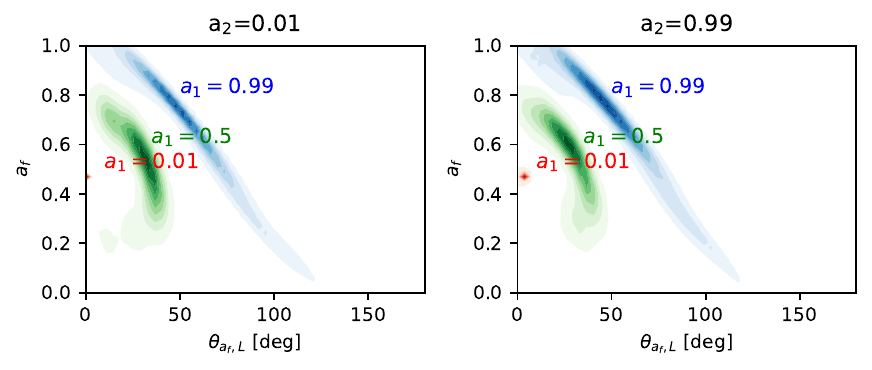}
\caption{{\bf Final spin magnitude and spin-orbital misalignment angle for 4:1 SMBH binary major mergers, differentiated by pre-merger spin vectors $\mathbf{a}_1$ and $\mathbf{a}_2$.} The color coding corresponds to different initial magnitudes of $\mathbf{a}_1$. The left panel displays scenarios where the secondary SMBH has a negligible initial spin, whereas the right panel represents cases with a nearly maximally spinning secondary SMBH. The pre-merger spins $\mathbf{a}_1$ and $\mathbf{a}_2$ are assumed to be isotropically distributed. }
\label{fig:merger-model}
\end{figure*}

\begin{figure*}
\centering
    \includegraphics[width=\columnwidth]{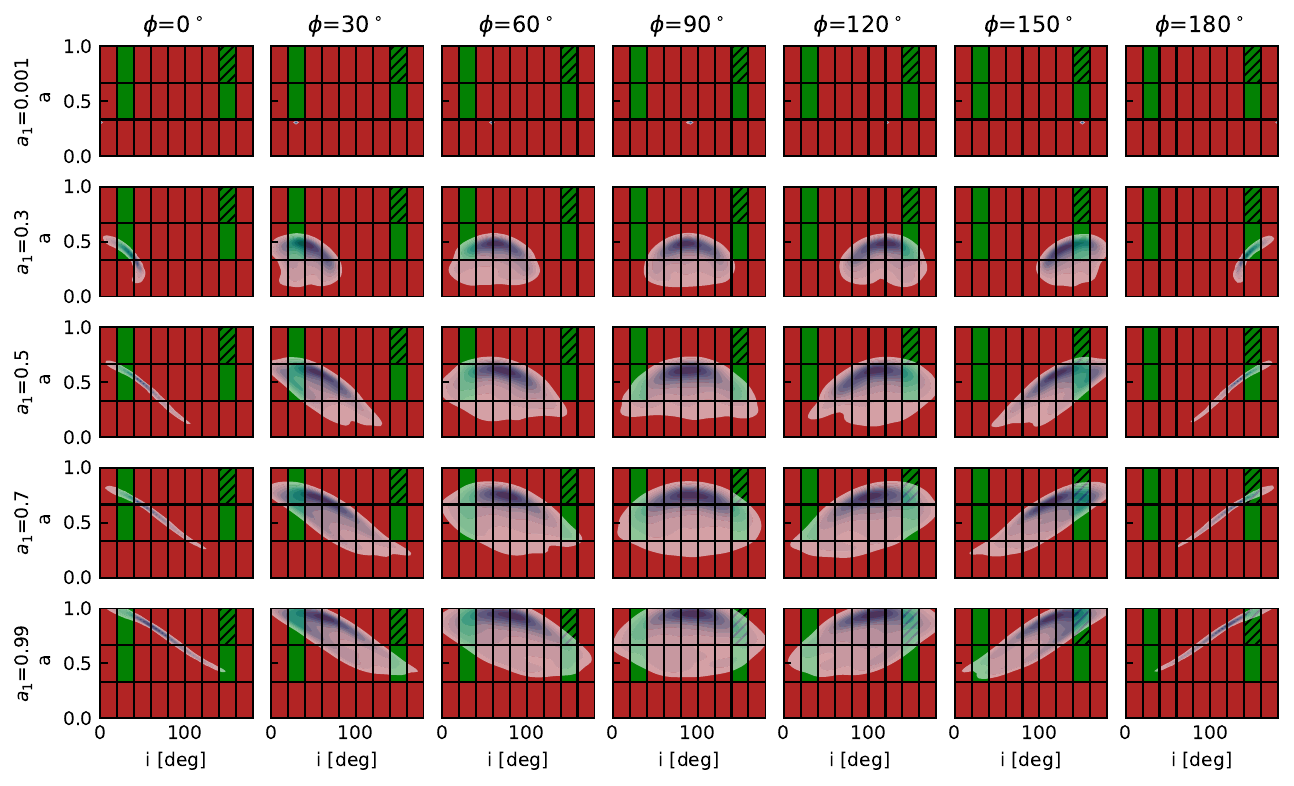}
    \caption{{\bf Kernel density estimates of the final BH spin and orientation from the 8:1 merger models.} $a_1$ indicates the spin magnitudes of the primary and secondary SMBHs before the merger, respectively. The value of $\phi$ represents different binary SMBH orientations with respect to the LOS. The orientations of $a_1$ are isotropically distributed to encompass both the accretion-only and merger progenitor cases. The final spin distributions show very weak dependency on $a_2$. }
    \label{fig:merger-stats-8}
\end{figure*}

\begin{figure*}
\centering
    \includegraphics[width=\columnwidth]{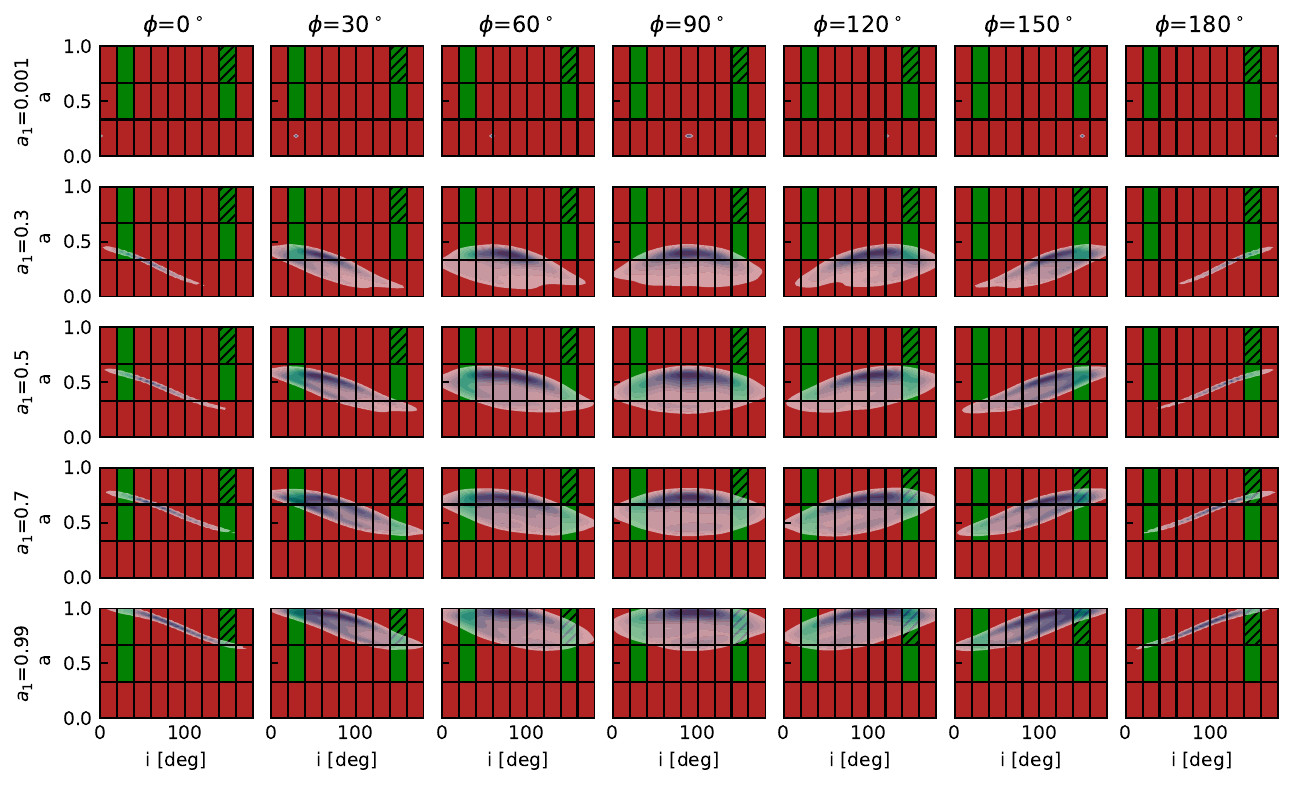}
    \caption{{\bf Kernel density estimates of the final BH spin and orientation from the 16:1 merger models.} $a_1$ indicate the spin magnitudes of the primary and secondary SMBHs before the merger, respectively. The value of $\phi$ represents different binary SMBH orientations with respect to the LOS. The orientations of $a_1$ are isotropically distributed to encompass both the accretion-only and merger progenitor cases. The final spin distributions show very weak dependency on $a_2$.}
    \label{fig:merger-stats-16}
\end{figure*}

\begin{figure*}
\centering
    \includegraphics[width=\columnwidth]{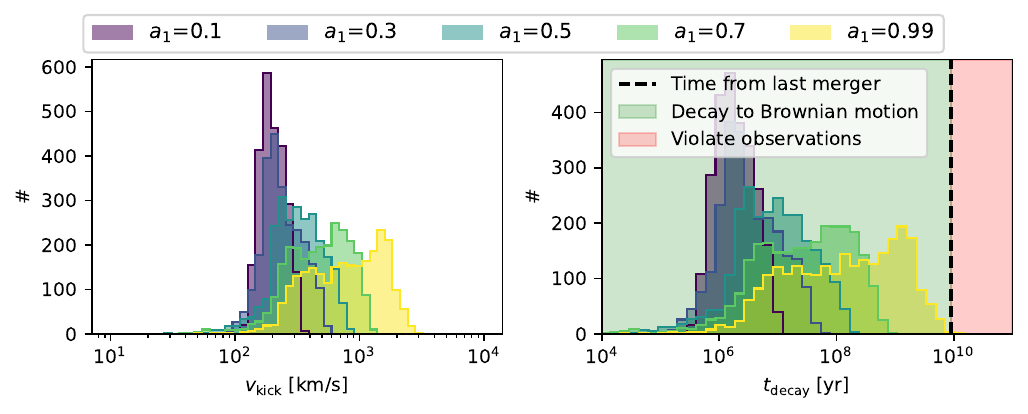}
    \caption{{\bf Velocity Distribution and Settling Time Post-4:1 Merger.} The left panel displays the distribution of the recoil kick velocities immediately following the 4:1 merger. The right panel depicts the required settling time for Sgr A* to return to the observed Brownian motion level at the Galactic Center.}
    \label{fig:merger-kick-4}
\end{figure*}

\clearpage

\end{document}